\documentclass[prl,aps,twocolumn]{revtex4}
\usepackage{graphics}
\begin{document}
\title{Field-tunable stochasticity in the  
magnetization reversal of
a cylindrical nanomagnet}
\author{Soumik Mukhopadhyay}
\author{Amrita Singh}
\author{Arindam Ghosh}
\affiliation{Department of Physics, Indian Institute of Science, Bangalore 560 012, India}

\begin{abstract}
The nature of magnetization reversal in an isolated cylindrical nanomagnet has
been studied employing time-resolved magnetoresistance measurement. We find
that the reversal mode is highly stochastic, occurring either by multimode or
single-step switching. Intriguingly, the stochasticity was found to
depend on the alignment of the driving magnetic field to the long axis of the
nanowires, where predominantly multimode switching gives way to single-step
switching behavior as the field direction is rotated from parallel to
transverse with respect to the nanowire axis. 
\end{abstract}


\maketitle

Traditionally, cylindrical nanomagnets have been of great interest for high
density magnetic storage~\cite{Nielsch_APL}, but very recently the
dynamics of domain walls (DW) in these systems is also rapidly gaining in
importance. A driving factor to this is the absence of ``Walker breakdown'',
with the DWs acting as ``massless'' particles having zero kinetic
energy~\cite{Ming_PRL}. However, a generic problem, in both planar or
cylindrical nanowires, is the stochasticity associated with the magnetization
reversal process. This is manifested in two classes of phenomena: first, the
stochasticity associated with diffusive and non-deterministic motion of the DWs
in presence of artificial or intrinsic disorder (addressed elsewhere by the
same authors~\cite{Singh_PRL}); second, the stochasticity related to the
nucleation and subsequent propagation of DWs. The latter has been investigated
in several planar magnetic films and lithographically patterned nanowires
through, for example, size and shape distributions of Barkhausen avalanches, or
extraordinary Hall effect etc ~\cite{Young_APL_PRL}.
Asymmetry in the mode of magnetization switching process on reversal of 
magnetic field polarity 
has been investigated using magnetic force microscopy~\cite{Suck_APL},
which might as well be related to the stochasticity issue (MFM probes only
a fraction of the whole sample). The stochasticity in
magnetization reversal in case of cylindrical nanowires, however, is relatively
unexplored.

\begin{figure}
\resizebox{8.5cm}{9.5cm}
{\includegraphics{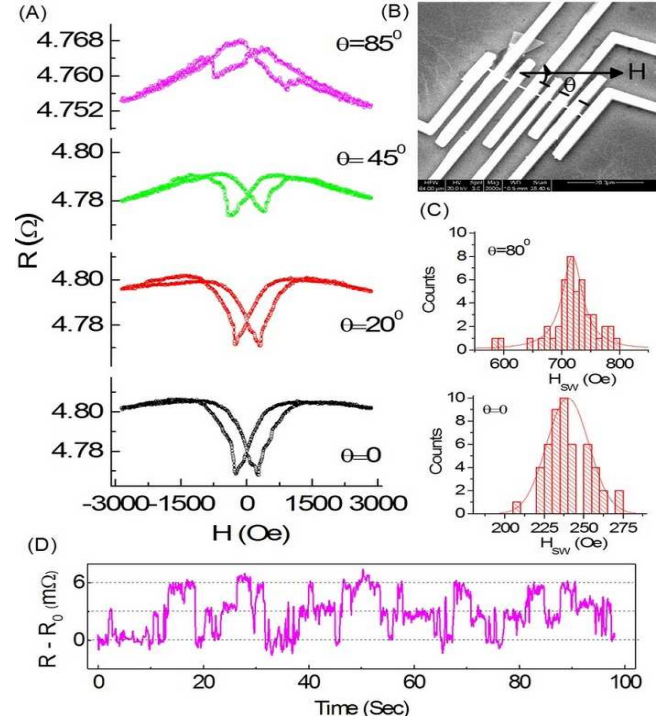}}
\caption{(A) Magnetic field dependence of resistance at different angles
between the applied magnetic field and the nanowire axis suggesting 
that the angle
dependence of $H_{sw}$ has $180^{\circ}$ periodicity. 
(B) SEM micrograph of a device. (C)
The switching field histograms for $\theta=0$ and $\theta=80^{\circ}$. 
(D) The time
dependence of resistance at a static low magnetic field ($1.5$ Oe) for
$\theta=0$ is shown.} \label{fig:mrtheta}
\end{figure}

Cylindrical magnets have been the classical template for theoretical study of
the mode of magnetization reversal and more importantly the angular variation
of the nucleation field, albeit in the limit of an infinite and isotropic
cylinder~\cite{Aharoni_PR}. Many experiments on
magnetization reversal in nanoscale magnetic systems have looked at the angular
dependence assuming an infinite and isotropic cylinder (or strips), although
finite size and anisotropy related effects can be crucial in those
cases~\cite{Wernsdorfer_PRL, Bryan_APL, Silva_PRB, Oliveira_PRB, Pignard_JAP,
Lavin_JAP, Rheem_JAPD, Rheem_Nanotech, Chung_JAP}. Wegrowe et al. could fit
their data on angular dependence with the curling mode prediction for an
infinite cylinder assuming an activation volume 
with aspect ratio $2:1$ (Ref.~\cite{Wegrowe_PRL}).
Moreover, surface anisotropy or structural defects can also play crucial role
and hence cannot be treated within the framework of isotropic
magnetization~\cite{Wegrowe_PRL}. The principal objective of this article is to
explore the effect of angular variation on the stochasticity of magnetization
reversal, rather than the variation in switching field ($H_{sw}$), where we
show that finite size, disorder, local anisotropy etc. play crucial roles.

Experimentally, we have taken a different approach. Instead of magnetic
hysteresis measurement such as using microfabricated
SQUID~\cite{Wernsdorfer_PRL} which is limited to low temperature, or
magneto-optic Kerr effect (MOKE)~\cite{Bryan_APL} which requires significant
averaging over many field cycles to improve the signal to noise ratio and hence
unsuitable for probing stochastic switching, we have measured time-resolved
electrical resistance around $H_{sw}$ to track the nucleation of individual DWs
at different angles between the applied magnetic field ($H$) and the long axis
of the cylindrical nanomagnet. Electrical transport in magnetic nano-cylinders
have been reported before~\cite{Wegrowe_PRL, Silva_PRB, Tanase_JAP}, 
but primarily to explore the
switching mechanism, rather than its stochasticity. Our approach and results
may also be relevant to device applications such as DW logic
systems~\cite{Allwood_JAP} or spiral turn sensors~\cite{Diegel_Sensorlett},
which require the use of orthogonal magnetic fields.

Nickel nanowires were electrochemically grown inside anodic alumina templates.
The average diameter of the nanowire is $\approx 200$~nm, where strong shape
anisotropy (aspect ratio $\sim 200$) aligns the magnetization along 
the long axis
of the nanowire. Details of the growth process and structural characterization
can be found elsewhere~\cite{Singh_APL_2009}. Following growth, nanowires were
dispersed on flat silicon oxide substrates, and electron-beam lithography was
used to form Ti/Au contact pads for electrical measurements on single nanowire
as shown in Fig.~\ref{fig:mrtheta}B. The resistance was measured at room
temperature in the
four-probe geometry using standard ac lock-in technique.

The magnetoresistance measured at different angles between the applied magnetic
field and the current or the long-axis of the cylinder show that the switching
field has a minimum at $\theta=0$ and a maximum at $\theta \sim \pm 90^{\circ}$
(Fig.~\ref{fig:mrtheta}A). This $180^{\circ}$ periodicity indicates curling-type
magnetization reversal mechanism expected for nanowires of diameter $200$ nm at
least when $H$ is parallel to the long axis~\cite{Aharoni_PR}. 
The $H_{sw}$ measurement was repeated $50$
times. At $\theta=0$, $H_{sw}$ has a distribution of $12\%$ around the most
probable value of $240$ Oe, while at $\theta=80^{\circ}$, the distribution is
slightly narrower, about $8\%$ with a most probable switching field of $720$ Oe
(Fig.~\ref{fig:mrtheta}C). Although such a narrowing for the transverse
direction has been observed before~\cite{Wernsdorfer_PRL} for Nickel nanowires
of diameter less than $100$ nm below $6$K, the overall width of
$H_{sw}$ distribution is significantly broader in our case, probably due to the
larger diameter of the nanowire (which could introduce inhomogeneties 
thereby lowering the local anisotropy barrier) and 
enhanced role of thermal activation at room 
temperature.

\begin{figure}
\resizebox{9cm}{13.5cm}
{\includegraphics{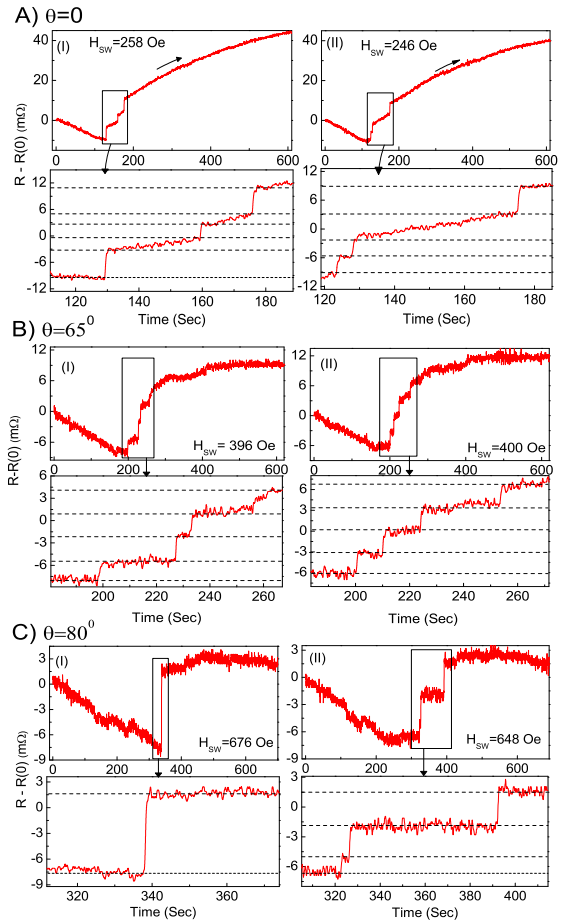}}
\caption{AI,II) Time dependence of resistance for two measurement cycles
with sweeping magnetic field at $\theta=0$ with $R(0)\sim 4.78$ m$\Omega$. 
The magnetization reversal
involves successive stochastic irreversible jumps in
resistance which are integer multiples of a $R_{dw}$. Between two jumps,
the resistance rises weakly due to AMR.
BI,II) Same for $\theta=65^{\circ}$. Resistive jump of size $2R_{dw}$ is rare
at high values of $\theta$.
CI, II) Time dependence of resistance
showing irreversible single-step and multimode switching
at $\theta=80^{\circ}$ for two measurement cycles. Absence of
AMR leads to flat resistance plateaus.
}\label{fig:MR0}
\end{figure}

For magnetoresistance to be a viable probe to magnetization reversal, it is
essential to identify and separate the contributions from anisotropic
magnetoresistance (AMR) arising due to spin-orbit scattering, and that from the
DWs. To achieve this we have studied the time dependence of nanowire resistance
($R$) at $H \ll H_{sw}$. The distance between the voltage 
probes placed midway 
along the length of the nanowire is 
$\sim 5\mu$m. The nanowire 
is magnetized in one direction along the long axis 
and then a small magnetic field
is applied in the opposite direction.
The time dependence of R at $H = 1.5$~Oe applied parallel to
the long-axis of the nanomagnet shows stochastic switching back and 
forth between
discrete multi-level states in $R$ (Fig:~\ref{fig:mrtheta}D). The jumps
describe increase in $R$ from its base value $R_0$ ($\sim 4.78 \Omega$) by
$\Delta R$ or $2\Delta R$, where $\Delta R \approx 3$~m$\Omega$ and vice versa.
Since $H$ is constant, AMR or Lorentz contributions to $R$ do not change, and
hence the observed switching between different R states can
be attributed to the DWs traveling in and out of the region between the voltage
probes. Increasing $R$ by $\Delta R$ and $2\Delta R$ corresponds to the
existence of one and two DWs, respectively, between the voltage probes. The
positive correction $\Delta R$ can be quantitatively understood from the
Levy-Zhang model of spin-mixing inside the DWs~\cite{Levy_PRL_1997} and
has been discussed in detail elsewhere~\cite{Singh_PRL}. 

\begin{figure}
\resizebox{8.5cm}{7.5cm}
{\includegraphics{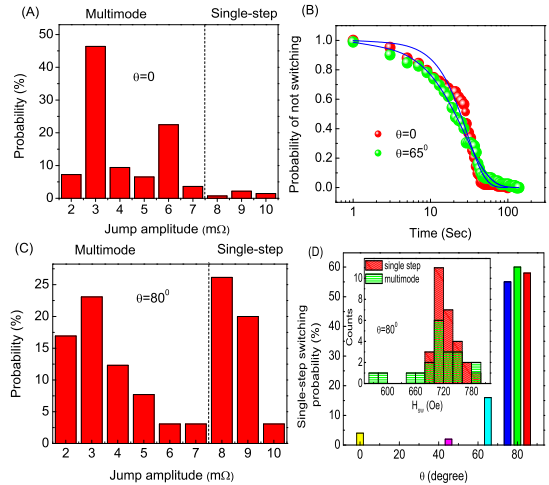}}
\caption{(A) The jump size distribution during the irreversible switching for
(A) $\theta=0$ (B) The enhanced-exponential probability of not
switching with $\beta>1$ ($\beta=1.86, 1.34$ for $\theta=0, 65^{\circ}$ 
respectively) indicates 
a relaxing multi-valleyed free energy landscape.
(C) Jump size distribution for $\theta=80^{\circ}$
(D) Probability of single-step switching
as a function of $\theta$. 
Inset: Switching field distribution for
single step and multimode reversal for $\theta=80^{\circ}$ is shown separately.}
\label{fig:stats}
\end{figure}

To study the dynamics of magnetization reversal, $H$ was swept at a rate of $2$
Oe/S across $H_{sw}$ while $R$ was measured as function of time 
with a resolution of $100$ mS.
Below $H_{sw}$, as $H$ is increased from zero after being saturated in a high
$H$ of opposite polarity, the resistance decreases monotonically. 
Above $H_{sw}$, successive
irreversible jumps in the resistance are observed at $\theta=0$
(Fig:~\ref{fig:MR0}A). Single-step switching is rare at $\theta=0$, and the
striking feature is that the most probable jump size is an integer multiple of
$3$~m$\Omega$ (Fig:~\ref{fig:stats}A). The time interval between two successive
jumps in a single cycle is non-deterministic and varies from one field cycle to
another (Fig.~\ref{fig:MR0}A, B). The resistance during that time interval is
not constant due to AMR and rises weakly with time (or $H$). When $H$ is
applied along the transverse direction single step switching events become more
frequent than multimode switching (Fig:~\ref{fig:MR0}CI) and the AMR not being
significant in transverse direction, $R$ stays constant between
successive jumps (Fig:~\ref{fig:MR0}CII).

The resistance jumps at integral multiple of $\Delta R$ allows us to understand
the magnetization reversal mechanism at $\theta = 0$. Since $\Delta R$
corresponds to the resistance of a single DW (vortex type since the $\theta$ 
dependence of $H_{sw}$ suggests curling mode reversal), it can be safely argued
that the stochastic multiple jumps in resistance correspond to multiple
nucleation events, where jumps of size $6$ m$\Omega$ indicate the nucleation of
a pair of DWs. Motion of DWs far below the nucleation field or magnetization
reversal due to nucleation of a DW pair is consistent with the time dependent
solution of Landau-Lifshitz equation~\cite{Broz_PRL} and particularly for
transition metal nanowires~\cite{Skomski_PRB}. 
The plateaus correspond to DW pinning
/ small-scale displacement or even rotation of spins 
(given the weak upward trend in the
resistance plateaus at $\theta=0$ due to AMR). Presence of multiple 
vortex DW suggests
that magnetization reversal at $\theta=0$ occurs through localized curling mode
with multiple nucleation events. Due to the simplicity of device geometry, 
it is an open question as to where the DW
nucleation starts (possibly at the wire ends or at some local inhomogeneties
within the wire) although they are eventually detected within the voltage
probes.  
Similar jumps in the
longitudinal resistance have been observed in FePd nanostructure 
~\cite{Danneau_PRL}. Multiple vortex walls have also been observed in
planar NiFe nanowire of larger width~\cite{Chung_JAP}.

The stochastic distribution of switching time between each irreversible jump
was measured over hundred such switching events. The switching probability was
computed from the integral over the switching time histograms for different
values of $\theta$ (where multimode switching is predominant), which were
fitted with a stretched exponential function $P(t)=\exp{-(t/\tau)^{\beta}}$,
with $\beta$ varying between $1.3-2$ (Fig:~\ref{fig:stats}B). The experimental 
data for $\theta=0$ follows the overall feature of the theoritical fit although
some deviation is observed for low time scale possibly due to small number
of switching events in that region.  
Similar results ($\beta > 1$)
have been reported elsewhere, which use more elaborate experimental techniques
such as micro-SQUID, magnetic force microscopy or
MOKE~\cite{Wernsdorfer_PRB_Co, Wernsdorfer_PRB, Lederman_PRL, Atkinson_NAT}.
The probability of not switching as shown in Fig.~\ref{fig:stats}B
characterized by $\beta > 1$ suggests reversal to involve correlated thermally
activated processes over a distribution of barrier heights in the free energy
landscape, rather than the Neel-Brown picture of thermal activation over a
single energy barrier. 

The distribution of jump size taken over $50$ measurement cycles at different
values of $\theta$ shows that for low angles such as $\theta=0$, the most
probable jumps are peaked around $3$ and $6$ m$\Omega$ while for higher angle
such as at $\theta=80^{\circ}$, an additional strong peak appears around $8-9$
m$\Omega$ with the distribution around $3$ m$\Omega$ being
significantly broadened (the resistive jumps of size $6$ m$\Omega$ are rare). 
It is readily observed that
in contrast to the preponderance of multimode switching at lower angle,
single-step switching is more frequent at higher angles 
(Fig:~\ref{fig:stats}C,D).

The origin of enhanced  
single-step switching at higher values of $\theta$ remains unclear.
Recent micromagnetic simulations have shown that application of transverse $H$ 
results in the expulsion of the vortex wall leading to
magnetization reversal via coherent
rotation~\cite{Silva_PRB}. 
However, the experimental evidence is merely based on
the deviation of $H_{sw}$ value from that predicted for curling mode at high
angle. No such
deviation has been observed for NiFe~\cite{Rheem_JAPD,Chung_JAP} or Ni
nanowires at low temperatures~\cite{Wernsdorfer_PRL}. Wegrowe et al. attributed
the deviation from curling mode for cylindrical Nickel nanowire (average
diameter $60$ nm) at room temperature to pinning by surface
defects~\cite{Wegrowe_PRL}. 
At higher $\theta$, the 
value of $H_{sw}$ predicted by coherent rotation model is 
smaller than that from curling
model~\cite{Pignard_JAP}.
In our case, the curling mode prediction for
a prolate spheroid~\cite{Aharoni_JAP} (assuming an exchange length $d_{ex}=40$
nm and bulk saturation magnetization $M_{S}=0.485$ T; the demagnetization
factors along the major and minor axes are $D_{z}=0.00015$ and $D_{x}=0.5$,
respectively) 
gives $H_{sw}=140$ Oe
at $\theta=0$. 
The underestimation of $H_{sw}$ value at $\theta=0$  
has been reported for nanowires with $d/d_{ex}>3$
~\cite{O'Barr_JAP}. The curling model overestimates $H_{sw}$ at 
$\theta=80^{\circ}$ by a 
factor of two, while, on the other hand, $H_{sw}$ value predicted by 
coherent rotation
model~\cite{Stoner_IEEE} (where $H_{sw}$ peaks at $\theta=0, \pm 90^{\circ}$) 
is nearly an order of
magnitude smaller compared to that observed experimentally at 
$\theta=80^{\circ}$. Moreover, we have observed that
single-step switching on the average involves higher $H_{sw}$
compared to multimode switching 
(Inset, Fig:~\ref{fig:stats}D). Therefore single step switching
cannot be associated with coherent rotation mode. 
One possible mechanism for single jump reversal is  
nucleation with very small activation volume and the
vortex wall sweeping across the entire length of the
nanowire~\cite{Hertel_PhysicaB}. The question is whether such a 
scenario is plausible 
in presence of transverse magnetic field.  
Transverse $H$
can modify pinning or depinning processes of DWs 
~\cite{Atkinson_APL}. 
Recently, huge
enhancement of DW velocity under strong transverse $H$ has been
reported~\cite{Richter_APL}.    
Nonetheless, our experiments prove the stochastic
nature of the mode of reversal at high values of $\theta$ with both single-step
and multiple-step reversal being possible.

The observation that application of transverse $H$ on a cylindrical nanomagnet
has a higher probability to avoid multimode switching could prove to be 
extremely important for device
applications. Large area nucleation pads are used for DW injection into a
nanowire where multimode switching due to transverse magnetic field has been a
major obstacle~\cite{Bryan_IEEE}. Bryan et al. observed multimode switching in
wide planar Permalloy nanowire, due to application of transverse
$H$~\cite{Bryan_APL}. Instead of planar wires of rectangular cross-section,
nucleation pads with square cross-section (to circumvent lithographic
obstacles in fabricating cylindrical nanowire) can be used to reproduce the
properties of a cylindrical nanomagnet. Use of softer magnets 
such as permalloy can reduce the switching field value in the transverse
direction.

To conclude, we have employed a simple time dependent magnetoresistance
measurement technique to probe the stochasticity of nucleation-mediated 
magnetization reversal in a
cylindrical nanomagnet. When the applied magnetic field is parallel to the long
axis of the nanomagnet, the magnetization reversal essentially follows the
localized curling mode with more than one nucleation events, whereas when the
magnetic field is applied perpendicular to the long axis of the nanomagnet, the
mode of reversal is strikingly non-deterministic and can either follow a
single-step switching or a
multimode switching process. 

We acknowledge the Department of Science and technology, Government of India
for funding the work. SM thanks CPDF, IISc for financial support.

\end{document}